\documentclass{vgtc}                          

\ifpdf
  \pdfoutput=1\relax                   
  \usepackage{graphicx}                
  \DeclareGraphicsExtensions{.pdf,.png,.jpg,.jpeg} 
\else
  \usepackage{graphicx}                
  \DeclareGraphicsExtensions{.eps}     
\fi%

\graphicspath{{figures/}{pictures/}{images/}{./}} 

\usepackage{microtype}                 
\PassOptionsToPackage{warn}{textcomp}  
\usepackage{textcomp}                  
\usepackage{mathptmx}                  
\usepackage{times}                     
\usepackage{cite}                      
\usepackage{tabu}                      
\usepackage{booktabs}                  
\usepackage{wrapfig}
\usepackage{url}
\usepackage{hyperref}
\usepackage[table]{xcolor}
\usepackage{pdfpages}

\newcommand{\etal}{et al.\@ }
\newcommand{\etals}{et al.\@'s }

\definecolor{linkColor}{HTML}{4682B4}
\newcommand{\mylink}[1]{{\color{linkColor} \texttt{#1}}}

\newcommand{\classWebsite}{\mylink{\href{https://www.mcnutt.in/dvpp-w21/}{mcnutt.in/dvpp-w21}}}
\newcommand{\paperWebsite}{\mylink{\href{https://www.mcnutt.in/zine-potential/}{mcnutt.in/zine-potential}}}

\newcommand{\midsepremove}{\aboverulesep = 0.0mm \belowrulesep = 0.0mm}

\newcommand{\parahead}[1]{\altparahead{#1.}}

\newcommand{\zine}[1]{\emph{#1}}
\newcommand{\numZines}{44}
\newcommand{\altparahead}[1]
{%
    \vspace{0.07in}%
    \noindent%
    \textbf{\textit{#1}}%
}
\newcommand{\secref}[1]{\hyperref[#1]{Sec.~\ref*{#1}}}
\newcommand{\noref}[1]{\hyperref[#1]{~\ref*{#1}}}
\newcommand{\appendixref}[1]{\hyperref[#1]{Appendix.~\ref*{#1}}}
\newcommand{\figref}[1]{\hyperref[#1]{Fig.~\ref*{#1}}}
\newcommand{\eqnref}[1]{\hyperref[#1]{Eqn.~\ref*{#1}}}
\newcommand{\tabref}[1]{\hyperref[#1]{Table ~\ref*{#1}}}

\onlineid{0}

\vgtccategory{Research}

\vgtcinsertpkg

\title{On the Potential of Zines as a Medium for Visualization}

\author{Andrew McNutt \thanks{e-mail: mcnutt@uchicago.edu}}
\affiliation{\scriptsize University of Chicago}

\abstract{
  Zines are a form of small-circulation self-produced publication often akin to a magazine.
  This free-form medium has a long history and has been used as means for personal or intimate expression, as a way for marginalized people to describe issues that are important to them, and as a venue for graphical experimentation.
  It would seem then that zines would make an ideal vehicle for the recent interest in applying feminist or humanist ideas to visualization. Yet, there has been little work combining visualization and zines.
  In this paper we explore the potential of this intersection by analyzing examples of zines that use data graphics and by describing the pedagogical value that they can have in a  visualization classroom. 
  In doing so, we argue that there are plentiful opportunities for visualization research and practice in this rich intersectional-medium.
} 

\CCScatlist{
  \CCScatTwelve{Human-centered computing}{Visu\-al\-iza\-tion}{~Visualization application domains}{}
}

\begin{document}


\firstsection{Introduction}

\maketitle


Visualizations are consumed in a vast array of mediums: ranging from the sprawl of digital options to printed materials to physicalizations.
Prior work has widely explored this space,  considering the roles that data graphics play in mediums as varied as watches\cite{blascheck2018glanceable} and gifs\cite{shu2020makes}.
Yet the modest form of the zine has received little attention as a venue for visualization research and practice.

\begin{figure*}[t]
  \centering
  \includegraphics[width=\linewidth]{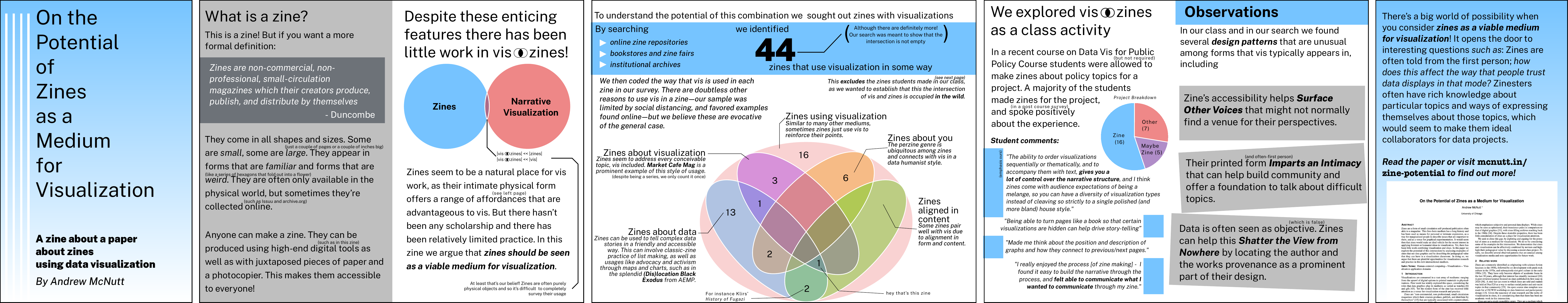}
  \caption{
    A zine summarizing the content of this paper.
    Zoom in to read or see the page-size version in the appendix.
  }
  \label{fig:full-zine-example}
  \vspace{-0.2in}
\end{figure*}

Zines are  ``non-commercial, non-professional, small-circulation magazines which their creators produce, publish, and distribute by themselves'' \cite{duncombe1997notes}
that are typically associated with counterculture, art, politics, and activism\cite{bigcartel}.
They come in many forms
(some are akin to artist's books\cite{thomas2009value} while others are closer to traditional books),
content (including collections of essays, comics, and genre-defying collages), and lengths (ranging from a couple of pages to hundreds).
Their typically physical nature offers an intimate and sustained way to engage with material that is often not present in digital media \cite{jabr2013reading} (although zines are sometimes distributed online in venues such as Issuu).
Some are regularly issued, while others are ad hoc.
Their ``do-it-yourself'' character imparts a rich accessibility,
as they can be cheaply produced using hand-drawn or collaged images (reproduced using a photocopier) as well as made with high-end desktop publishing systems\cite{wan1999not}.
They are produced for many reasons, including personal expression, teaching aids, technical explanation, or sometimes the output of workshops \cite{struzek2021designing}.
An example of a zine can be found in \figref{fig:full-zine-example}.

This array of affordances and complications would seem to make zines an enticing area of consideration for visualization.
Their graphically laden form matches well with the types of spaces that visualizations can occupy.
The low barrier to entry presents opportunities for the inclusion of voices often left out of more traditional publishing\cite{gisonny2006zines} such as queer people \cite{creasap2014zine}, people of color\cite{bold2017diverse}, and incarcerated people\cite{duncombe1997notes}---which aligns with D'Ignazio and Klein's call for feminist data practices\cite{d2020data}.
The often hand-made and personal nature of many zines maps on to Lupi's Data Humanism\cite{lupi2017data}, which emphasizes subjective and personal data displays.
While zines may be seen as ephemeral, their transience pales in comparison to that of digital graphics \cite{kosara2016Bits}, with zines filling archives reaching back to the 1960s\cite{hays2020citation}.
Despite these desirable properties, there has been little consideration of zines as a place for visualization attention.

We seek to close this gap, by exploring and arguing for the potential of zines as a medium for visualization.
We do so by considering some of the examples in this intersection.
We demonstrate that zines and visualization can be effectively combined by novices and highlight their pedagogical value by discussing a recent class project.
Finally, we describe several observed patterns that are unusual among visualization media and note opportunities for future work.

\section{Related work}

Zines are commonly identified as originating with science-fiction fanzines in the 1930s, followed by co-development with punk-rock culture in the 1970s, and subsequently riot grrrl culture in the early 1990s \cite{duncombe1997notes}.
They have only become objects of academic focus in the last 30 years, although that interest has steadily increased\cite{hays2020citation}.
A peer-reviewed journal focused on zines published its first issue in 2020\cite{Zines20Etienne}.
A zine fair (an event in which zines are sold and traded) was held at FAccT20 as a way to surface social justice and anti-racist topics in that community\cite{hanna2020ctrlz}.
An open source zine-template was made for a CSCW20 workshop on data feminism and participatory design\cite{DIgnazio2020ZineTemplate}.
Given the nascency of zine research and the rarity of visualization in zines,  it is unsurprising then that there has been no academic work in this intersection.

Zines are not a single monolithic genre.
They are a medium which can contain countless narrative forms.
Duncombe describes a zine taxonomy consisting of political, personal, scene, network, fringe culture, religious, vocational, health, sex, travel, literary, and art zines, as well as comix and ``the rest''\cite{duncombe1997notes}.
New forms and genres of zine continue to develop today \cite{bigcartel}.
For instance, the genre of tech zines---which describe topics such as
quantum computing\cite{franklin2020},
tcp, git \cite{evansWizardZines}, and AI\cite{MotherPeople}---has grown in popularity in recent years, with dedicated zine fairs emerging to support them \cite{nycTechZine, tinyTechZines}.
The popularity of this genre may be due to the often casual form of zines, allowing for gentle introductions to technical or daunting topics \cite{wibowo18Zines}.
This penchant has been seen previously, such as in the Judith Butler fanzine \zine{Judy!}\cite{lorusso_2019}.
They are closely related to data comics\cite{bach2018design, zhao2015data}, a genre of narrative visualization which communicates data-rich topics through comics.
E.g. similar to Wang \etals\cite{wang2020data} data comics about user studies,
the  \zine{Denizen Designer}\cite{DenizenDesigner} zine uses data graphics to describe a survey of community-based designers and introduce participatory design.
While zines resemble well studied visualization media, their traditions, tendencies, and free-form nature differentiate them from previously explored forms like posters or postcards.

While there are some zines explicitly about visualization, we do not seek to insert a notion of visualization zine into Duncombe's taxonomy (or to Segel and Heer's narrative visualization genres \cite{segel2010narrative}) as such a structure would have unusefully porous boundaries---reflecting the amorphous nature of zines themselves\cite{thomas2009value}.
What makes zines worthy of study in visualization is not their potential for a single mode of expression, but instead the tensions of an ambiguously-defined often-physical medium that has a tradition of personal rhetoric, which is sometimes understood to be cheap\cite{hays2020citation}
and is accessible to marginalized voices\cite{citeThis, gisonny2006zines}.

\section{Visualization in Zines in Practice}\label{sec:theming}

Compared to the hundreds of thousands of extant zines\cite{duncombe1997notes}, there are a relatively limited number of zines that use visualization.
This may be because, as Duncombe argues, that zines tend to favor personal narratives and stories and not facts and figures \cite{duncombe1997notes}. This would suggest then that the language of personal expression found in zines is poorly matched with the regimentation and abstraction typical to visualizations.
However, we argue that this notion overly simplifies what zines can and have been, as well as what purposes visualization can serve therein.
For instance, Poletti\cite{poletti2008popular} notes that list making is a common practice in zines that serves various purposes, such as ``in personal zines, lists commonly disrupt linear narrative and are often used to evoke very specific moments''.
This is similar to the role the charts play in other mediums\cite{segel2010narrative}.
Further: lists---like spreadsheets or tables---can be seen as a simple form of visualization, as they transform data into a visual form interpretable by a human. For instance the NYT's front-page list of nearly 100k COVID-19 victims \cite{Grippe20} communicates the magnitude of the tragedy in a tangible and graphical manner.

Further countering the idea that visualizations are not at home in zines is the existence of examples of this intersection.
We now explore a series of representative instances organized into usage themes.
The intent of this sample is not to provide a complete survey, but instead to demonstrate that this intersection is non-empty.
We developed these themes by gathering a sample from as many resources as were reasonably available during COVID-19-era social distancing. This involved contacting zine librarians and fairs, visiting zine bookstores, and trawling online zine repositories and stores.
Our sample yielded \numZines{} zines using visualization in some way (see appendix or \paperWebsite{}).
We open coded them with an interest in determining the usages and purposes that visualization serves in this medium.
This yielded 5 highly overlapping categories (\figref{fig:full-zine-example}).
While this categorization is simple, it highlights the types of tasks that charts perform in zines from a high-level.
There may be other purposes that might have served, however our sample was limited in scope, considered a medium with a long history, and was biased towards zines available online---limitations which may have affected the types of zines we identified.

\begin{figure*}[t]
  \centering
  \includegraphics[width=\linewidth]{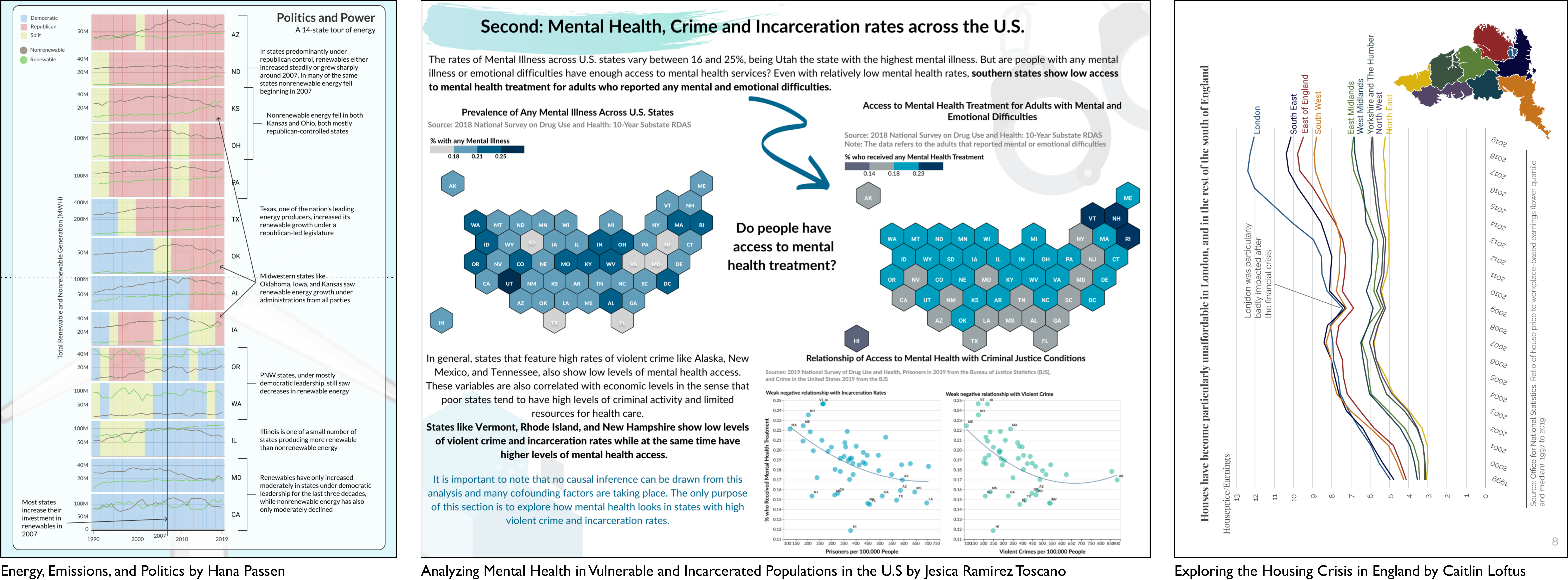}
  \caption{
    Excerpts from zines submitted to the static project in our data visualization for public policy course, reproduced with permission.
    Being unbound from the compositional restrictions of forms like web pages prompted exploration and creativity in annotation, encoding, and layout.
  }
  \label{fig:class-examples}
  \vspace{-0.2in}
\end{figure*}

\parahead{Zines about data}
Despite common assertions to the contrary; the meaning of data is variable, subjective, and human\cite{gitelman2013raw,correll2019ethical,d2020data}.
Zines offer an interesting position from which to interrogate the view-from-nowhere gauze of objectivity that is often inhabited through the use of data \cite{d2020data} and charts\cite{mcnutt2020surfacing}, as subjectively-inflected narratives are so common in that medium.
The roles and rhetorical positions that utilizing zines as means for explaining data can vary, including using it as a mechanism for advocacy, for documentation, or for easing technical complexity through gentle explanations.
For instance, \zine{(Dis)location Black Exodus}\cite{dislocationblackexodus} is a print/online zine produced by Anti-Eviction Mapping Project that highlights the unequal levels of eviction and displacement brought on by gentrification in San Francisco. It does so by weaving together oral histories with maps and figures produced by members of the communities affected by this displacement.
D'Ignazio and Klein\cite{d2020data} highlight that many of these maps shirk so-called best-practices or notions of clean-design to drive home the message that ``there are too many evictions''. They highlight that what might be understood as just a dot on a map, is a real person who has been evicted.
The presentation of these figures in a zine lends them a physicality and a sense of volume that might not be available in a webpage---again underscoring the message that there are too many evictions and that each eviction is personal.

Zine's proclivity for (sometimes encyclopedic) lists
meshes well with displays of data.
\zine{Mapping Out Utopia}\cite{devin17Mapping} explores counter-culture in 1970s Boston through a directory-style presentation of counter-culture groups---such as feminist bookstores and alternative newspapers. Each entry describes the group, their address, years of operation, and the current value and use of that physical location; a juxtaposition meant to highlight the area's gentrification and changing culture. These lists are supplemented with maps that, like \zine{(Dis)location}, are less concerned with visual clarity and adherence to notions of ``graphical excellence'', and more with expressions of place, context, and volume.
Verbose data representations with summative review have a history in zines, for instance Amateur Press Associations (a form of collective epistolary zine) are sometimes referred  to as ``data zines''\cite{duncombe1997notes}.

Just as tech zines use the medium as a way to give easy introduction to complex ideas, zines about data can help gently guide the reader through unfamiliar datasets and topics.
Zhao's \zine{Garbage}\cite{GarbageZhao} utilizes elements of data-comics
to highlight ecological costs of waste in the United States. Some pages feature a named and drawn anthropomorphic character learning about the effects waste has on the environment, while others have more conservative presentations of data through compositions of maps and charts.
The seamless transition between each of the seemingly conflicting modes of expression highlights malleability of the form.

\parahead{Zines about visualization}
Given the universe of topics that zines address, it is natural that there would be some zines that are about visualization itself.
\zine{Market Cafe Mag}\cite{marketCafeMag} is a regularly produced high production zine that focuses on design and cultural aspects of visualization and surrounding topics, with issues specifically dedicated to topics such as trusting data and time.
Each issue features a variety of types of content including articles on the issues topic, documentation of projects, and interviews (which are a zine staple \cite{duncombe1997notes}).
Watson's \zine{Visualization Magazine}\cite{VisualisationWatson} ran for several issues during the late 2000s and featured full page reproductions of then-contemporary infographics, essays, and book reviews from a variety of authors.
This bricolage is evocative of the form of many zines: they contain a variety of content which is assembled to paint a picture of a topic.
Zines can be used to teach complex information and ideas, of which visualization certainly can be a topic.

\parahead{Zines using visualization}
Beyond their unique features, zines---like other mediums---sometimes use data graphics as a component of their production. Just as a book or magazine may feature a chart, some zines use charts to serve a brief rhetorical purpose.
\zine{Paper Cuts Library \#2}\cite{paperCutsLibrary} exemplifies such a strategy as it explores the causes, contributors, and impacts of the recently ended ``longest bull market in United States history.'' The majority of work is text describing its subject, as might be found in a typical magazine article, punctuated by occasional line charts, bar charts, and big number displays whose purpose is to underscore and enhance the narrative, rather than focusing on presence or form of a visualization itself.
\zine{Biff Boff Bam Sock \#8}\cite{Beck18Biff} provides an overview of how to navigate American health insurance system through a first person narrative that is interspersed with light or comedic asides to provide respite from potentially dry material. It uses tables and charts to illustrate its points, such as a bar chart indicating that the price of insurance is correlated with age.
While this usage may only differ marginally from other media, its context differs as many zines use first person narrative (often explicitly situating the authors), casting such graphics in a different rhetorical light than the disembodied perspective found in many presentations.

\parahead{Zines about you}
Zines allow for personal expression in a manner that most other mediums do not. This is likely  due to history with individualistic or alternative cultures \cite{duncombe1997notes}, as well as the amorphous nature of their form---typically any approach to their design or content is valid. The freedom of personal expression is thus not bound by the restrictive mores found in other forms.
For instance, \zine{``I can't explain it any other way''}\cite{lilianZine}
documents a person's concert attendance over the preceding year---evidently a cornerstone of her identity---and pairs various maps and charts with lyrics that have emotional resonance for her.
Especially poignant is the post-text listing her intended concert attendance for the summer of 2020, which was disrupted by the COVID-19 pandemic.
It's charts and maps are, or appear to be, hand drawn or sketchy style, yielding a Dear Data-style \cite{lupi2016dear} intimacy in the form, further underscoring the sense that the reader is participating in something personal.
Adorning the cover are memes, highlighting the free-wheeling visual-recycling nature of the medium, as well as its allowance for using humor.
While the success of Dear Data has shown that expressive humanist graphics can be of value, there have been relatively few mediums that support such work---beyond the postcards exemplified in that project. Zines can fill this gap by offering a flexible format that itself has a close connection with personal expression.

\parahead{Aligned with the form} The content of a piece can sometimes motivate and inform the selection of its medium.  The fanzine \zine{Visualizing the History of Fugazi }\cite{KlirsFugazi} describes the history of the American post-hardcore band Fugazi through infographics and maps.
It is heavily design focused pages include displays such as ``The International Fugazi Tour Network'' (a connected arc-map of performance locations), and ``Fugazi's Fellow Travelers'' (an arc diagram mixed with a time-series of bands with whom they frequently performed).
This set of displays could have been prepared as a series of posters or a webpage, however fanzines are closely associated with the culture surrounding bands like Fugazi, and so the form taken on by this project mirrors the culture of the content it describes.
Zines can be a useful medium for data displays if their content is aligned in some way with the medium, such as in the Fugazi zine, which describes an aspect of a culture associated with zines, or through reflexivity, as in our zine describing this paper in \figref{fig:full-zine-example}.

\section{Visualization in Zines in Pedagogy}\label{sec:pedagogy}

Zines have long been used as a teaching tool\cite{wan1999not} for a variety of topics, including composition, gender studies, math, creativity, and the arts \cite{Byers20You, hays2020citation}.
Teaching zine-making, Creasap argues, exemplifies feminist principles, such as ``participatory learning, validation of personal experience, and the development of critical thinking skills"~\cite{creasap2014zine}.
Zines have value in those contexts because they foster a personal engagement with the material.
We argue that the zine's pedagogical utility may have a place in visualization classrooms, and in doing so underscore the accessibility of the medium and it's pliability to realistic visualization problems.

We examined this assertion in a recent iteration of a course on data visualization for public policy.
The course is centered around a term-long project wherein students select a policy topic of interest for which they build a static data narrative and an interactive web-based visualization.
The static half of the course supported Altair\cite{vanderplas2018altair} (although any Python library was allowed) and Figma as a layout tool.
Students were in their second year of a computational public policy masters program  and had at least a year of Python and statistical experience with most having little experience with visualization.

In this iteration, the static component was required to be printable, and zines were noted as a possible form (as were data comics, posters, and reports), however their use was not required or incentivized.
$16/28$ chose to produce works that could easily described as zines (with 5 that might arguably classified in that manner)---each of which fell into our \textbf{\emph{zines about data}} theme.
\figref{fig:class-examples} excerpts several submitted zines.
Some students elected to have their work shown on the course page (\classWebsite{}).
Topics included workplace gender inequality, the relationship between renewable energy and political polarization, and mental health in American prisons.
One used a data comic-style anthropomorphic narrator to guide the reader through an analysis of early childhood development programs in Peru.
Students seemed enthusiastic about the medium, many of whom expressed enthusiasm for the assignment in a post-course survey.
One student noted ``I really enjoyed the process -  I found it easy to build the narrative through the process, and felt able to communicate what I wanted''; however, the need to join several tools sometimes impeded this fluidity.
We were generally impressed by the quality of the produced work, which demonstrated good understanding of sequencing, narrative, and composition.
While policy students are unlikely to make zines in their professional lives (a limitation of this project), we believe that they will likely have constraints imposed on their work and will need to be creative in their data presentations, activities for which making zines prepares them.

We believe that zines are an enticing medium for teaching visualization, as they offer a low barrier to entry and promote compositional thinking.
Zine's free-form format seemed to empower students, yielding richer and more nuanced compositions than observed in previous iterations of the course.
Similarly to how Bach \etal \cite{bach2018design} argue that teaching data comics promote compositional thinking, we suggest that zines may provide an even more accessible means to similar ends, as they require less artistic skill to produce.
Beyond this pedagogical value, the success of these projects shows that zines and visualization can be combined effectively by novices with relatively little training; underscoring the low barrier to entry.








\section{Notable patterns}

We observed several notable patterns in our student's work and our survey.
While not every zine exhibited all (or any) of these patterns, we note them because they are unusual among visualization media.

\parahead{Surfacing Other Voices}
Anyone of any technical skill level can make a zine. It is a medium that has traditions but not rules, and is thus open and accessible to all\cite{duncombe1997notes}.
This is a pattern that we also see in zines that use visualization: they offer a home for presentations of information from voices that might not make it to privileged visualization venues.
The maps and histories found in \zine{(Dis)Location}\cite{dislocationblackexodus} or the encyclopedic documentation of \zine{Mapping out Utopia}\cite{devin17Mapping} are unlikely to have appeared in another form of publication.
The freedom of form yields a freedom of expression: it can be used to convey the perspectives, concerns, and values of marginalized and underrepresented people \cite{gisonny2006zines, citeThis}.

\parahead{Imparting an Intimacy} Zines are often framed as first person narratives, and those that feature visualizations are no different. The effect that this has varies, but it often creates a space of intimacy between the author and the reader\cite{duncombe1997notes}---a closeness whose origin extends beyond zines' sometimes hand-drawn form.
This intimacy can be used to discuss personal or difficult topics---for instance, a student of Cresap's describes ``his experiences with coming out as a young trans man'' through a zine\cite{creasap2014zine}.
A review of McNeil's chart-rich personal zine notes that it displayed ``a level of vulnerability that I wouldn’t have thought possible through charts and graphs''\cite{seagreenzines}.
One of our students noted that zine making ``felt more personal than pretty much any other visualizations I made.''
In addition to vulnerability or candor, this intimacy can be conveyed by humor, such as in \zine{Biff Boff Bam Sock}\cite{Beck18Biff}, which can in turn be used to soften the impact of dry or technical material,
as well as to consolidate an in-group or build community. For instance, \zine{Two-Fisted Library Studies}\cite{fisted17} uses charts and graphs to make jokes about the experience of being a librarian,
and in doing so, provide a point of connection for those in the library community.
These intimacies may favorably impact engagement, memorability, and perhaps even empathy (an emotion that may be out of reach for visualization \cite{correll2019ethical}).

\parahead{Shattering the View from Nowhere} Many of the zines examined directly locate the author relative to the work. For instance, a number of zines we have looked at in this paper \cite{paperCutsLibrary, KlirsFugazi, devin17Mapping} explicitly highlight the subjective nature that their authors played in creating them, and describe the limitations of their work and their perspective.
For instance Klirs notes  ``Any data visualization project involves some subjective choices''~\cite{KlirsFugazi}.
Dislocation of the false-objectively that often enshrouds data and data visualizations\cite{mcnutt2020surfacing} is one of the key advantages of this form---the typically personalized nature of the medium is beneficial not just for its intimacy, but because by being personal it locates the author within the work, and therein serves goals of disclosure and transparency.
This agrees with prior arguments\cite{correll2019ethical, d2020data} that provenance must play a larger role in visualization---that the positions and biases of the authors and data should be described so as to contextualize and situate the design.

\section{Discussion}\label{sec:discussion}

Progenitor of zine scholarship, Duncombe, has been paraphrased as saying ``Zines have already been discovered, and it's up to us to do something interesting with them'' \cite{wan1999not}. Visualization has been and will be used in zines,
so it is up to us to consider the research questions that this intersection prompts.

\altparahead{What can visualization do for zines?}
Zines are lauded for having a low barrier to entry\cite{citeThis}, as anyone with access to a photocopier can produce them, however zines are also produced using high-end desktop publishing.
Just as tools like Idyll \cite{conlen2018idyll} support the medium of interactive articles, future work might consider how tool design can support zine-making tasks for users without limited technical expertise.
For instance, the ``scruffy'' zine aesthetic\cite{duncombe1997notes} can be difficult to reproduce with high end tools.
Our students used an assortment of tools to produce their zines whose disconnection created undue repeated work with one student noting that they kept
``locking the design of something before decid[ing] to change it again, so it was a little hard to iterate''---a problem that Bigelow \cite{bigelow2016iterating} also observed in artists.
Such tool research could expand on tools that use hand-drawn aesthetics\cite{xia2018dataink, kim2019dataselfie} by better tuning them to zines users and uses (such as visual-recycling or scruffy aesthetics), or Electric Zine Maker~\cite{LawheadElectric}, which is a playful tool with intentional connections to glitch art for creating zines, but lacks links to data.

In our usage theming (\secref{sec:theming}) we considered the purposes that visualizations can have in zines, although future work might expand on this by considering their rhetorical roles.
The examples considered used a wide variety of chart forms, ranging from simple graphics, such as bar charts, to unusual or ad hoc graphical forms.
This suggests that zines may be pliable to experimentation with unusual data presentations.

\altparahead{What can zines do for visualization?}
Zines raise questions about data storytelling and the rhetorical tactics\cite{hullman2011visualization} used therein. For example: what effect does embracing the personal affectation have on data comprehension? How do zine's physical-yet-ephemeral nature modify their experienced meaning? How does their non-objective form influence presentation of uncertainty or affect reader trust?
As zines have a history of fostering community\cite{duncombe1997notes} any research or tool building should be done in partnership with the people who might be affected by the results.
Just as Li \etal~\cite{li2021we} urge researchers to view artists as viable technical collaborators, so too do we highlight zinesters as being potentially valuable partners in understanding this form and its consequences for visualization.
In this work we have seen that zines offer a useful venue for explaining data, but they can also be used by researchers to explain and promote their work\cite{fox2016extensions}
(as in \figref{fig:full-zine-example}).
Their concrete physicality can offer a respite from the vastness\cite{creasap2014zine} of the always omnipresent online panopticon\cite{bigcartel}.
Zines and visualization have shown to be a useful combination:
multiple visualization zines have won design awards \cite{KlirsFugazi, marketCafeMag},
a wide space of examples have occurred naturally,
and it appears to have value in the visualization classroom.

We thus believe that zines have a rich potential as a medium for visualization research and practice.

\acknowledgments{We thank our reviewers, André G. Wenzel, Quimby's Bookstore, Ravi Chugh, Katy Koenig, and Lilian Huang.}

\bibliographystyle{abbrv-doi}

\bibliography{zine-paper}

\clearpage
\pagebreak
\appendix{}
\section{Supplementary Materials}

In this appendix we provide additional details about various aspects of the work that were unable to fit in the main paper.

\subsection{Survey}

To better understand the student experience of the zine project during our course we conducted a post-course survey. We draw quotes from this survey in the main body of the text. This Google Forms-hosted survey  was conducted 2 months after the course.
Students who participated were paid $\$6$.
A reproduction of the study instrument can be found below. Unless otherwise noted each of the responses was configured to be a paragraph style input box. The survey was conducted non-anonymously (in order to verify payment information), and then was anonymized prior to analysis.

\begin{enumerate}
  \item Prior to the course how familiar were you with zines?

        \begin{tabular}{lllllll}
          Completely & 1       & 2       & 3       & 4       & 5       & Completely \\
          Unfamiliar & $\circ$ & $\circ$ & $\circ$ & $\circ$ & $\circ$ & Familiar
        \end{tabular}

  \item Prior to the course how familiar were you with visualization?

        \begin{tabular}{lllllll}
          Completely & 1       & 2       & 3       & 4       & 5       & Completely \\
          Unfamiliar & $\circ$ & $\circ$ & $\circ$ & $\circ$ & $\circ$ & Familiar
        \end{tabular}
  \item Did you produce a zine for your submission?

        \begin{tabular}{l}
          $\circ$ Yes                                                    \\
          $\circ$ My submission could have been debatably called a zine. \\
          $\circ$ No                                                     \\
          $\circ$ I'm not sure                                           \\
        \end{tabular}
  \item If you chose to make a zine, why did you pick that format? If not, why not? Did you have prior zine-making experience?
  \item Considering both your own work and that of your classmates, do you think zines are a useful medium for visualization? If so, what aspects of zines are conducive to forming visualization narratives? If not, why not?
  \item Compare the process of making your printable submission to independent or group projects in other classes. Did you find it more engaging or feel a greater sense of ownership than with other similar projects? Did the process of preparing a printable object change your relationship with your topic?
  \item Which tools did you use to create your submission? Please describe your experience using these tools. What was difficult? What was easy? Are there any features you wish existed as part of this combination of tools?
  \item If you produced a zine, did you enjoy making the zine making process? Do you think you will make zines again? Do you believe that the process of making this zine will have value for your personal development or your career?
  \item Is there anything else you wish to share with us?
\end{enumerate}

Out of the 28 students who took the class, a total of 7 students responded to the survey.
Thus, we do not report  aggregate statistics as they would only reflect those who elected to participate, and not necessarily the character of the class as a whole.

We sought permission from several students to include their projects in this work (specifically in \figref{fig:class-examples}). Students who allowed us to reproduce their work in this paper were paid $\$25$.

\pagebreak

\subsection{Zine Survey Listing}

In \tabref{tab:zine-list} we list the zines that appeared in our survey.
This selection of zines was found by consulting zine librarians, zine bookstores, online archives (such as Issuu and archive.org), and institutional zine archives.
There are, of course, additional zines that would be appropriate to consider in this survey, however we did not encounter them.
There were a number of zines featuring visualizations that we became aware of but were not able to obtain or view, and hence do not appear in this listing.
The intent of this survey was to establish the existence of examples of zines using visualization that have come about naturally.
Thus while a total survey of zines using visualization is out of scope of this work (and given the physical nature of zines, possible anywork), we are able to demonstrate that examples do in fact exist.

\subsection{Example Zine}

We include a full page version of the example zine shown in \figref{fig:full-zine-example} on the final page of this supplement, such that interested parties can print it for themselves, and thus be provided an example of a zine using visualization. It follows standard conventions for constructing a single-page single-sided 8 page zine. Instructions for preparation can be found in various locations online, such as those found at \url{https://www.wikihow.com/Make-a-Zine}.

\clearpage
\pagebreak

\newcommand*\rot{\rotatebox{90}}

\begin{table*}
\caption{A listing of the zines that appeared in our sample, along with how they were categorized in our usage theming. Terse names are used in the interest of presentation, but they correspond to the usage themes listed in \secref{sec:theming}.}
\label{tab:zine-list}
\midsepremove
\begin{tabular}{rp{4.5in}lllll}
&
Zine name and citation &  
\rot{Uses vis.} & 
\rot{About  data}& 
\rot{About  vis.} & 
\rot{About  you} & 
\rot{Aligned in form}
\\
\toprule\\
& Counts & 
44 & 
18 & 
5 & 
8 & 
3
\\
\toprule\\
1 & Aarati Akkapeddi. \emph{Encoding}. Undated & \cellcolor{cyan!25} Yes & \cellcolor{cyan!25} Yes & \cellcolor{red!30} No & \cellcolor{cyan!25} Yes & \cellcolor{red!30} No\\ 
2 & Andrew McNutt. \emph{Design and analysis of table cartograms}. 2019 & \cellcolor{cyan!25} Yes & \cellcolor{red!30} No & \cellcolor{cyan!25} Yes & \cellcolor{red!30} No & \cellcolor{red!30} No\\ 
3 & Andrew McNutt. \emph{On the Potential of Zines as a Medium for Visualization}. 2021 & \cellcolor{cyan!25} Yes & \cellcolor{red!30} No & \cellcolor{cyan!25} Yes & \cellcolor{red!30} No & \cellcolor{cyan!25} Yes\\ 
4 & Anna Jo Beck. \emph{Biff Boff Bam Sock \#8: For your health }. 2019 & \cellcolor{cyan!25} Yes & \cellcolor{red!30} No & \cellcolor{red!30} No & \cellcolor{red!30} No & \cellcolor{red!30} No\\ 
5 & Annie Schleser. \emph{perfect!}. 2009 & \cellcolor{cyan!25} Yes & \cellcolor{red!30} No & \cellcolor{red!30} No & \cellcolor{red!30} No & \cellcolor{red!30} No\\ 
6 & Anti-eviction mapping project. \emph{(Dis)location Black Exodus}. 2019 & \cellcolor{cyan!25} Yes & \cellcolor{cyan!25} Yes & \cellcolor{red!30} No & \cellcolor{red!30} No & \cellcolor{red!30} No\\ 
7 & Brett Bloom / BKDN BKDN. \emph{BREAK DOWN WORKBOOK \#2—SONIC MEDITATIONS: INVESTIGATING PETRO-SUBJECTIVITY}. 2019 & \cellcolor{cyan!25} Yes & \cellcolor{red!30} No & \cellcolor{red!30} No & \cellcolor{red!30} No & \cellcolor{red!30} No\\ 
8 & Bubblesort Zines. \emph{Hip-Hip Array}. 2016 & \cellcolor{cyan!25} Yes & \cellcolor{red!30} No & \cellcolor{red!30} No & \cellcolor{red!30} No & \cellcolor{red!30} No\\ 
9 & Carni Klirs. \emph{Visualizing the History of Fugazi}. 2018 & \cellcolor{cyan!25} Yes & \cellcolor{cyan!25} Yes & \cellcolor{red!30} No & \cellcolor{red!30} No & \cellcolor{cyan!25} Yes\\ 
10 & Cecilia Zhao. \emph{GARBAGE}. 2019 & \cellcolor{cyan!25} Yes & \cellcolor{cyan!25} Yes & \cellcolor{red!30} No & \cellcolor{red!30} No & \cellcolor{red!30} No\\ 
11 & Charlotte Francoeur, Lindsey Leigh. \emph{The Wonderful World of Microbes Written}. 2021 & \cellcolor{cyan!25} Yes & \cellcolor{cyan!25} Yes & \cellcolor{red!30} No & \cellcolor{red!30} No & \cellcolor{red!30} No\\ 
12 & Corinne Halbert, Erika Iris and Demi Zigler. \emph{Mathematics for Misfits}. 2021 & \cellcolor{cyan!25} Yes & \cellcolor{red!30} No & \cellcolor{red!30} No & \cellcolor{red!30} No & \cellcolor{red!30} No\\ 
13 & Equity and Health Innovations Design Research Lab. \emph{The Denizen Designer Zine}. 2021 & \cellcolor{cyan!25} Yes & \cellcolor{cyan!25} Yes & \cellcolor{red!30} No & \cellcolor{red!30} No & \cellcolor{red!30} No\\ 
14 & FAQNP: A Queer Nerd Publication. \emph{Queer Nerds Visualized}. 2014 & \cellcolor{cyan!25} Yes & \cellcolor{red!30} No & \cellcolor{red!30} No & \cellcolor{cyan!25} Yes & \cellcolor{red!30} No\\ 
15 & Gabe Wigtil. \emph{The Solar System to Scale - science astronomy mini zine}. Undated & \cellcolor{cyan!25} Yes & \cellcolor{red!30} No & \cellcolor{red!30} No & \cellcolor{red!30} No & \cellcolor{red!30} No\\ 
16 & Hannah Israelsohn. \emph{Language Zine}. 2008 & \cellcolor{cyan!25} Yes & \cellcolor{red!30} No & \cellcolor{red!30} No & \cellcolor{red!30} No & \cellcolor{red!30} No\\ 
17 & Joe Song. \emph{Awesome Attributes About Ants}. 2009 & \cellcolor{cyan!25} Yes & \cellcolor{red!30} No & \cellcolor{red!30} No & \cellcolor{red!30} No & \cellcolor{red!30} No\\ 
18 & Joe Song. \emph{The beginner's guide to variable stars }. Undated & \cellcolor{cyan!25} Yes & \cellcolor{red!30} No & \cellcolor{red!30} No & \cellcolor{red!30} No & \cellcolor{red!30} No\\ 
19 & Joe Song. \emph{The beginner's guide to binary stars }. Undated & \cellcolor{cyan!25} Yes & \cellcolor{red!30} No & \cellcolor{red!30} No & \cellcolor{red!30} No & \cellcolor{red!30} No\\ 
20 & Johnny Masiuleqicz. \emph{Happy Tapir \#3 A cup of holiday fear}. 2017 & \cellcolor{cyan!25} Yes & \cellcolor{red!30} No & \cellcolor{red!30} No & \cellcolor{cyan!25} Yes & \cellcolor{red!30} No\\ 
21 & Jupiter Hadley and Dann Sullivan. \emph{Seeds}. 2016- & \cellcolor{cyan!25} Yes & \cellcolor{cyan!25} Yes & \cellcolor{red!30} No & \cellcolor{red!30} No & \cellcolor{red!30} No\\ 
22 & Lilian Huang. \emph{I can't explain it any other way': a zine of Lilian's concert attendance, 2019}. 2020 & \cellcolor{cyan!25} Yes & \cellcolor{cyan!25} Yes & \cellcolor{red!30} No & \cellcolor{cyan!25} Yes & \cellcolor{red!30} No\\ 
23 & Lily Xie. \emph{Charts}. Undated & \cellcolor{cyan!25} Yes & \cellcolor{red!30} No & \cellcolor{red!30} No & \cellcolor{cyan!25} Yes & \cellcolor{red!30} No\\ 
24 & Liz Mason. \emph{Most Unwanted Zine}. 2021 & \cellcolor{cyan!25} Yes & \cellcolor{cyan!25} Yes & \cellcolor{red!30} No & \cellcolor{red!30} No & \cellcolor{cyan!25} Yes\\ 
25 & luxoft. \emph{Tech Spark}. 2021 & \cellcolor{cyan!25} Yes & \cellcolor{red!30} No & \cellcolor{cyan!25} Yes & \cellcolor{red!30} No & \cellcolor{red!30} No\\ 
26 & Melissa Simerson. \emph{Superfoods}. Undated & \cellcolor{cyan!25} Yes & \cellcolor{cyan!25} Yes & \cellcolor{red!30} No & \cellcolor{red!30} No & \cellcolor{red!30} No\\ 
27 & Papercuts library. \emph{\#2: Results of the longest bull market in american history}. 2020 & \cellcolor{cyan!25} Yes & \cellcolor{cyan!25} Yes & \cellcolor{red!30} No & \cellcolor{red!30} No & \cellcolor{red!30} No\\ 
28 & Patrice Solomon, Jarad Solomon. \emph{A science of mind and non-physical realms}. 2019 & \cellcolor{cyan!25} Yes & \cellcolor{red!30} No & \cellcolor{red!30} No & \cellcolor{red!30} No & \cellcolor{red!30} No\\ 
29 & Pocket Lint. \emph{Pocket Lint's Party Zine Spring 2018}. 2018 & \cellcolor{cyan!25} Yes & \cellcolor{red!30} No & \cellcolor{red!30} No & \cellcolor{red!30} No & \cellcolor{red!30} No\\ 
30 & Pocket Lint. \emph{Crying on Campus}. Undated & \cellcolor{cyan!25} Yes & \cellcolor{cyan!25} Yes & \cellcolor{red!30} No & \cellcolor{red!30} No & \cellcolor{red!30} No\\ 
31 & Rolling Thunder: An Anarchist Journal of Dangerous Living. \emph{Rolling Thunder 10 (Summer 2012)}. 2012 & \cellcolor{cyan!25} Yes & \cellcolor{red!30} No & \cellcolor{red!30} No & \cellcolor{red!30} No & \cellcolor{red!30} No\\ 
32 & Sarah McNeil. \emph{Statistical Things That Happen But Don’t Matter}. 2011 & \cellcolor{cyan!25} Yes & \cellcolor{red!30} No & \cellcolor{red!30} No & \cellcolor{cyan!25} Yes & \cellcolor{red!30} No\\ 
33 & Sarah McNeil. \emph{Updated Report of Observations Which Somehow Seem to Reflect Abstractly on Life}. 2011 & \cellcolor{cyan!25} Yes & \cellcolor{red!30} No & \cellcolor{red!30} No & \cellcolor{cyan!25} Yes & \cellcolor{red!30} No\\ 
34 & Sarah Poppy Jackson. \emph{Reclaim your sh*t!}. 2019 & \cellcolor{cyan!25} Yes & \cellcolor{cyan!25} Yes & \cellcolor{red!30} No & \cellcolor{red!30} No & \cellcolor{red!30} No\\ 
35 & The Sophie Lab. \emph{Newcomb tech in mind: the zine of of newcomb's institute of technology}. 2018- & \cellcolor{cyan!25} Yes & \cellcolor{cyan!25} Yes & \cellcolor{cyan!25} Yes & \cellcolor{red!30} No & \cellcolor{red!30} No\\ 
36 & Tim Devin. \emph{Mapping out utopia: 1970s Boston-area counterculture}. 2017 & \cellcolor{cyan!25} Yes & \cellcolor{cyan!25} Yes & \cellcolor{red!30} No & \cellcolor{red!30} No & \cellcolor{red!30} No\\ 
37 & Tina Richardson. \emph{Stepz: a psychogeography and urban aesthetics zine}. 2015 & \cellcolor{cyan!25} Yes & \cellcolor{red!30} No & \cellcolor{red!30} No & \cellcolor{red!30} No & \cellcolor{red!30} No\\ 
38 & Tiziana Alocci, Piero Zagami. \emph{Market cafe magazine}. 2017 - & \cellcolor{cyan!25} Yes & \cellcolor{red!30} No & \cellcolor{cyan!25} Yes & \cellcolor{red!30} No & \cellcolor{red!30} No\\ 
39 & Tom Vague. \emph{Vague 28: Wild West 11}. 1997 & \cellcolor{cyan!25} Yes & \cellcolor{red!30} No & \cellcolor{red!30} No & \cellcolor{red!30} No & \cellcolor{red!30} No\\ 
40 & Two Photon Art. \emph{Nicotine Zine}. 2019 & \cellcolor{cyan!25} Yes & \cellcolor{cyan!25} Yes & \cellcolor{red!30} No & \cellcolor{red!30} No & \cellcolor{red!30} No\\ 
41 & Two Photon Art. \emph{The Opium Poppy}. 2016 & \cellcolor{cyan!25} Yes & \cellcolor{cyan!25} Yes & \cellcolor{red!30} No & \cellcolor{red!30} No & \cellcolor{red!30} No\\ 
42 & Two-Fisted Library Studies. \emph{Volume 8, Number 1}. 2017 & \cellcolor{cyan!25} Yes & \cellcolor{red!30} No & \cellcolor{red!30} No & \cellcolor{red!30} No & \cellcolor{red!30} No\\ 
43 & Unknown. \emph{Fuck solitary: abolition \& HB5417}. Undated & \cellcolor{cyan!25} Yes & \cellcolor{red!30} No & \cellcolor{red!30} No & \cellcolor{cyan!25} Yes & \cellcolor{red!30} No\\ 
44 & Victoria Harley, Katrina Eresman. \emph{Terrific trees of the USA!}. Undated & \cellcolor{cyan!25} Yes & \cellcolor{cyan!25} Yes & \cellcolor{red!30} No & \cellcolor{red!30} No & \cellcolor{red!30} No
\end{tabular}
\end{table*}

\clearpage
\pagebreak
\includepdf{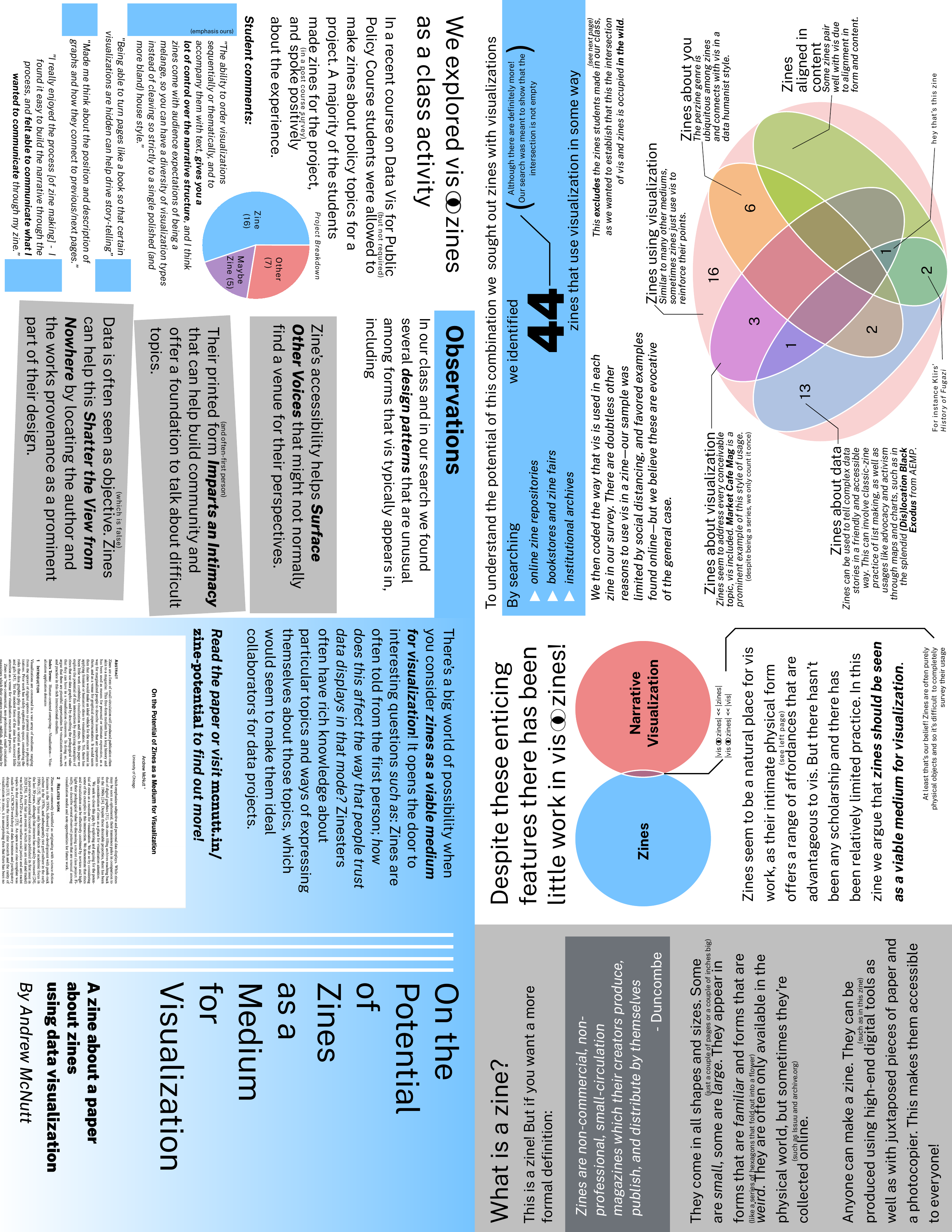}

\end{document}